\documentclass[12pt,english]{article}
\usepackage{mathptmx}
\usepackage[scaled=0.85]{helvet}

\usepackage[T1]{fontenc}
\usepackage[latin9]{inputenc}
\usepackage[letterpaper]{geometry}
\usepackage{float}
\usepackage{amsmath}
\usepackage{graphicx}
\usepackage{setspace}
\usepackage{amssymb}
\makeatletter
\usepackage{type1cm}\renewcommand{\normalsize}{%
   \@setfontsize\normalsize{12pt}{13pt}
   \abovedisplayskip 9\p@ \@plus2\p@ \@minus5\p@
   \abovedisplayshortskip \z@ \@plus3\p@
   \belowdisplayshortskip 6\p@ \@plus3\p@ \@minus3\p@
   \belowdisplayskip \abovedisplayskip
   \let\@listi\@listI}\normalsize\renewcommand{\small}{%
   \@setfontsize\small{10.5pt}{11.5pt}
}\small\makeatother

\usepackage{babel}

\makeatother\usepackage{upgreek}

\addto\captionsenglish{\renewcommand{\refname}{}}\usepackage{bibspacing}

\DeclareMathSizes{10}{9}{7}{6}

\usepackage[font={small},labelfont=bf,margin=0pt,belowskip=0pt]{caption}\usepackage[comma,authoryear]{natbib}
\bibpunct{(}{)}{;}{a}{}{;}
\renewcommand{\cite}{\citep}

\def\@biblabel#1{[#1]}
\newcommand{\bbar}{\ensuremath b\kern-0.7em-}
\newcommand{\bbars}{\ensuremath b\kern-0.6em-}
\newcommand{\lbbar}{\ensuremath b\kern-0.43em-}
\usepackage[normalem]{ulem}
\makeatother

\begin{document}
\textbf{\Large The end of nanochannels}{\Large \par}

\textit{Thomas B. Sisan}\footnotemark,\footnotetext{Dept of Physics,
Northwestern University, Evanston, IL 60208, USA} \textit{Seth Lichter}\setcounter{footnote}{0}\global\long\def\thefootnote{\fnsymbol{footnote}}
 \footnotemark\footnotetext{Corresponding author, Dept of Mechanical
Engineering, Northwestern University, E-mail: s-lichter@northwestern.edu,
\mbox{Phone: 847-467-1885}}

\noindent \vspace{-0.1in}

\begin{spacing}{1.5}
\noindent \textbf{Abstract:} Current theories of nanochannel flow
impose no upper bound on flow rates, and predict friction through
nanochannels can be vanishingly small. We reassess neglecting channel
entry effects in extremely long channels and find violations at the
nanoscale. Even in frictionless nanochannels, end effects provide
a finite amount of friction. Hence, the speed at which nanochannels
transport liquids is limited. Flow-rate and slip length measurements
are reevaluated using calculations which include end-effect friction.
End effects are critical for the design of new technological devices
and to understand biological transport.
\end{spacing}

\begin{spacing}{2.0}
\end{spacing}

\vspace{0.2in}

\noindent \setlength{\intextsep}{1cm plus1cm minus1cm}Nanoscale channels,
such as aquaporin and carbon nanotubes, exhibit surprisingly large
flow rates \cite{Preston1992,hummer_water_2001,kalra_osmotic_2003,falk_molecular_2010,thomas_reassessing_2008}.
Filtration membranes fabricated from carbon nanotubes have been reported
with flow rates 100 - 100,000 times greater than have been measured
for any other material \cite{majumder_nanoscale_2005,majumder_new_2011,holt_fast_2006,Dai_superlong_2011}.
Arbitrarily high measured flow rates are fit to theoretical calculations
by choosing sufficiently small values for the friction parameter along
the channel length \cite{majumder_nanoscale_2005,holt_fast_2006,Dai_superlong_2011}.
These calculations, however, ignore \textit{end effects}, viscous
losses within the liquid near the channel's entrance and exit. For
nanotubes whose typical lengths are 1000's times their radii, classical
fluid dynamics predicts that the great length of the channel provides
the primary flow resistance and thus, end effects are negligible.
The narrow and tortuous channel of aquaporin leads to similar expectations
of negligible end effects. However, contrary to usual fluid mechanics
practice and expectation, our results show that long low-friction
channels and short channels of any channel friction can have significant
end effects. Thus, all treatments of nanochannel flow must be aware
of end effects. Friction due to end effects is essential for calculating
physically realistic values for flow rates in nanochannels such as
carbon nanotubes and aquaporin. Our modified Hagen-Poiseuille equation
(see Methods) gives predictions comparable to molecular dynamics results
\cite{kalra_osmotic_2003,suk_water_2010,Thomas2009,Goldsmith2009}.
In addition, our results provide a theoretical limit to flow rates
in any nanochannel, and serve as an upper bound for experimental measurements.%
\begin{figure}
\begin{centering}
\includegraphics[scale=0.9]{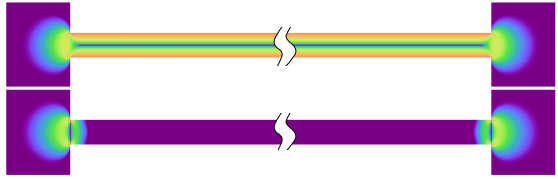}\caption{\textbf{Figure 1. }Energy dissipation (high (red), low (purple)),
represented by a perceptually balanced color scale (see Methods),
as numerically computed for a cylindrical channel between two large
reservoirs (2-D cross-section, only a small portion of the reservoirs
is shown). For a traditional high-friction {}``no-slip'' channel
(top panel), energy dissipation occurs predominantly along the channel
length. For a zero-friction channel (bottom panel), energy dissipation
occurs solely near the entrance and exit of the channel, irrespective
of channel length, in regions of high streamline curvature.}

\par\end{centering}

\end{figure}

For short aquaporin or slippery carbon nanotube channels, total resistance
within the channel can be surprisingly small. Previous efforts to
compute flow rates through these channels have focused exclusively
on the friction within the channel. But, regions of high viscous shear
persist near the channel ends, where streamlines sharply curve from
the large reservoirs into (and out of) the small lumen of the channel,
see \textbf{Figure 1} \cite{sampson_stokess_1891,weissberg_end_1962,suk_water_2010}.
The size of these localized regions is small, on the order of the
channel radius, yet including them in theoretical calculations shows
their contribution to flow resistance can be significant. In the limit
of zero channel friction $(f=0)$, flow resistance arises solely through
shear in these regions near the channel ends, see Figure 1, bottom
panel, thus creating a fundamental flow-rate speed limit, in contrast
to prior theory.

\textbf{Figure 2} shows experimental measurements (symbols) of flow
rate in channels of various lengths, $L$. %
\begin{figure}[H]
\begin{centering}
\includegraphics{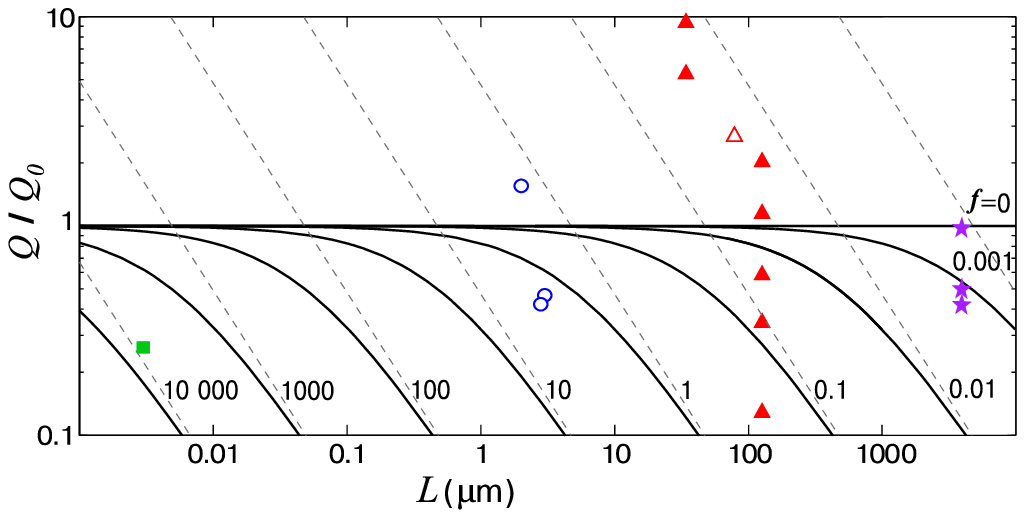}
\par\end{centering}

\caption{\textbf{Figure 2. }A flow rate speed limit in nanochannels. Predictions
(solid lines) of flow rate $Q$ with channel friction $f$ relative
to the frictionless-channel flow rate $Q_{\mathit{0}}$, versus channel
length $L$, are compared with measurements of carbon nanotubes (blue
circles \cite{holt_fast_2006}; red triangles, filled \cite{majumder_nanoscale_2005}
and open \cite{majumder_new_2011}; purple stars \cite{Dai_superlong_2011})
and aquaporin (green square \cite{Preston1992}). Several reported
flow rates \cite{majumder_nanoscale_2005,majumder_new_2011,holt_fast_2006}
are larger than allowed for frictionless channels, $f=0$. The dashed
lines show predictions, for each value of $f$, when ignoring end
effects \cite{majumder_nanoscale_2005,holt_fast_2006,Dai_superlong_2011}.
These lines erroneously suggest that any $Q$, no matter how large,
is theoretically possible. }

\end{figure}

\noindent Predictions that ignore end effects (dashed lines) increase
without bound, so a value of the friction factor $f$ can be found
to fit any empirical measurement \cite{majumder_nanoscale_2005,holt_fast_2006,Dai_superlong_2011}.
Predictions that include end effects (solid lines) asymptote to a
maximum flow rate, $Q_{0}$. The nine measurements that fall below
this line $(f=0)$ comply with known physics of fluid flow, while
the six above this line are unphysically fast. For measurements below
the limit, the theory predicts new lower values of $f$, and thus
higher values of the \textit{slip length} (see Methods). For example,
new predictions for the two overlapping blue circles, see Figure 2,
give values for the friction factor (slip length) that are half (twice)
as large as previously found. For flow measurements closer to the
limit $Q_{0}$, the discrepancy between old and new predictions diverges.

In summary, as friction within channels decreases, end effects make
an increasingly large contribution to total flow resistance, see \textbf{Figure
3}, even in extremely long channels where end effects have been rotely
neglected. For short channels, such as aquaporin, end effects remain
significant, asymptoting to a finite value, regardless of the amount
of channel friction, see Figure 3, green line. In previous calculations,
which omitted end effects, flow rates were unbounded as $f\rightarrow0$.
Since viscous losses at the channel entry and exit are independent
of channel friction, including them produces an upper limit to the
flow rate (Figure 2). Prior measurements that show flow rates above
this limit may point to difficulties in accurately determining channel
radii, net flow rates, or the number of channels spanning the membrane.
Errors may also arise due to compromise in membrane integrity. Appreciation
of end losses is necessary for understanding the high, but not unphysically
high, flow rates found in carbon nanotubes, so as to ultimately realize
the far-reaching %
\begin{figure}[H]
\noindent \centering{}\includegraphics{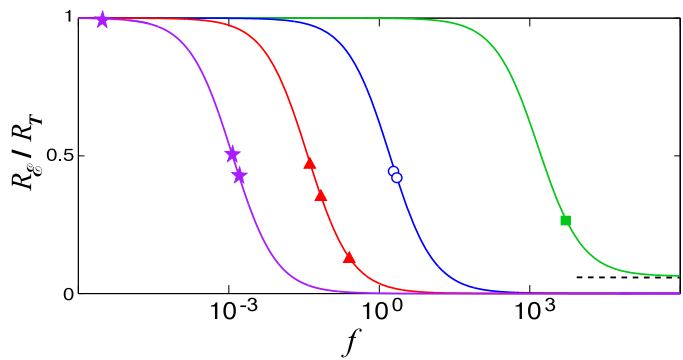}\caption{\textbf{Figure 3. }Low friction channels are dominated by end effects.
As channel friction $f$ decreases, flow resistance $R_{\mathcal{E}}$
due to viscous friction at the entrance and exit becomes a larger
fraction of the total flow resistance $R_{T}$. The colored lines
use the channel length, channel radii, pressure drop, and fluid viscosity
of the experimental measurements using Equations \ref{eq:endpressure}
and \ref{eq:modHP} (see Methods). The symbols (same as in Figure
2) correspond to those data in Figure 2 below the line $f=0$, and
are placed here by solving Equation \ref{eq:slipeq} (see Methods).
Surprisingly, extremely long channels (purple, red, and blue data)
can show significant end effects. Limit of green line as $f\rightarrow\infty$
is $\sim R/L$ (dashed line).}

\end{figure}

\noindent applications of fluid-transporting nanochannels. Our results
also indicate that end effect friction in biological channel entryways
may be a significant component in determining and controlling flow
rates \cite{hummer_water_2001}.

\vspace{16pt}

\noindent \textbf{Discussion:} The validity of our continuum scale
predictions have yet to be carefully tested for the narrowest channels
where molecular discreteness effects may be prominent. The published
flow rates from molecular dynamics studies \cite{kalra_osmotic_2003,suk_water_2010,thomas_reassessing_2008,Goldsmith2009}
are close to the frictionless channel limit. This is in contrast to
experimental reports which show some flow rates significantly above
this limit, see Figure 2. However, the molecular dynamics studies
do point to areas of further investigation. The pressure drop at the
entrance and exit of a channel are predicted by the continuum model
to be equal, but in \citet{suk_water_2010}, are not. This may indeed
be a phenomenon unique to molecular scale flows. However, it may also
be a result of the small size of the fluid reservoirs (the continuum
model assumes infinite reservoirs), to the unique way in which the
upstream and downstream reservoirs are connected, to non steady-state,
or even to nonhomogeneous system heating.

\noindent \vspace{16pt}

\noindent \textbf{Methods:} The Hagen-Poiseuille equation gives the
pressure drop, neglecting end effects, for fully-developed flow through
a cylindrical channel, \begin{equation}
\Delta P_{HP}=\frac{8\mu LQ}{\pi r^{4}+4\pi r^{3}/f},\label{eq:HPeq}\end{equation}
 where $Q$ is flow rate, $\mu$ is dynamic viscosity, $r$ is channel
radius, $L$ is channel length, and the friction factor \[
f=1/b,\]
where the slip length $b$ quantifies the amount by which the liquid
slips along the solid channel wall \cite{majumder_nanoscale_2005,holt_fast_2006}.
$Q$ diverges to infinity as $f\rightarrow0$.

The pressure drop due to end effects, $\Delta P_{\mathcal{E}}$, is
well approximated by the expression for flow through an aperture \cite{sampson_stokess_1891,weissberg_end_1962,suk_water_2010},
\begin{equation}
\Delta P_{\mathcal{E}}\approx\frac{3Q\mu}{r^{3}}.\label{eq:endpressure}\end{equation}
 Using the expression for total pressure drop across a channel between
two reservoirs, $\Delta P_{T}=\Delta P_{\mathcal{E}}+\Delta P_{HP}$,
Equations \ref{eq:HPeq} and \ref{eq:endpressure} give a modified
Hagen-Poiseuille equation, \begin{equation}
\Delta P_{T}=\left[\frac{3\mu}{r^{3}}+\frac{8\mu L}{\pi r^{4}+4\pi r^{3}b}\right]Q_{T}\label{eq:modHP}\end{equation}
where the term in brackets $\left[\ldots\right]$ is defined as the
total flow resistance $R_{T}$, see Figure 3. Equation \ref{eq:modHP}
can be rearranged to give the flow rate in the limit of large slip
$(b\gg r)$, \begin{equation}
Q=Q_{0}\left[1+\frac{2}{3\pi}\frac{L}{b}\right]^{-1},\label{eq:flowrate}\end{equation}
 where $Q_{0}=\Delta P_{T}r^{3}/3\mu$ is the flow rate through a
frictionless channel $(b\rightarrow\infty)$, and the subscript $T$
on the flow rate has been dropped for simplicity. Equation \ref{eq:flowrate}
takes into account both high slip and end effects in nanoscale channels,
in contrast to the normal HP equation (\ref{eq:HPeq}). By equating
the flow rate without and with end effects (Eqs. 1 and 3), the slip
length can be expressed in terms of previously-found slip lengths
$\bbars$ which neglected end effects, $b=\bbar(1-\frac{3\pi\lbbar}{2L})^{-1}$.
Experimental measurements which found $\bbar/L>2/3\pi$ are unphysical.

Fabricated carbon nanotube membranes \cite{majumder_nanoscale_2005,holt_fast_2006,hinds_aligned_2004,Dai_superlong_2011}
are comprised of a heterogenous distribution of nanotube radii, so
the total membrane flow rate is \[
Q_{M}=\sum_{i=1}^{N}Q\left(r_{i}\right),\]
where $N$ is the number of open nanotubes spanning the membrane and
$Q(r_{i})$ is the flow rate through channel $i$. We generate a normal
distribution of radii, $r_{M}$, based on published radii histograms
\cite{holt_fast_2006,hinds_aligned_2004,Dai_superlong_2011}. From
this distribution the flow rate through a membrane comprised of frictionless
carbon nanotubes is

\vspace{-0.2in}
\[
Q_{M,0}=\sum_{i=1}^{N}Q_{0}\left(r_{i}\right).\]
 The slip length can then be determined from Eq. \ref{eq:flowrate},
\begin{equation}
b=L\frac{2}{3\pi}\frac{\mathcal{Q}}{1-\mathcal{Q}},\label{eq:slipeq}\end{equation}
 where $\mathcal{Q}=Q_{M}/Q_{M,0}$ is the ratio of the measured flow
rate to the maximum rate that would occur through the same distribution
of frictionless nanotubes. %
\begin{table}[!]
\centering{}\includegraphics{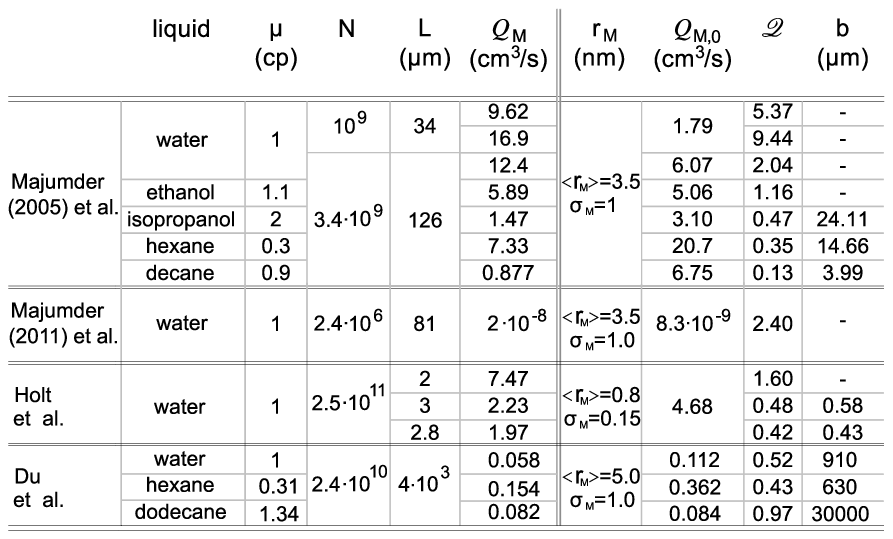}\caption{\textbf{Table 1. }Data from prior work \cite{majumder_nanoscale_2005,majumder_new_2011,holt_fast_2006,Dai_superlong_2011}
(left of vertical dividing line) and derived values (right of dividing
line) used in Figure 1. $\mu$ is dynamic viscosity, $N$ number of
open nanotubes spanning $1\,\mathrm{\mathsf{cm^{2}}}$ of membrane,
$L$ nanotube length, and $Q_{M}$ total flow rate through $1\,\mathrm{\mathsf{cm^{2}}}$
membrane. For \citet{majumder_nanoscale_2005, majumder_new_2011},
$Q_{M}$ was computed from their published velocities. For \citet{holt_fast_2006},
$Q_{M}$ was computed from their published slip lengths. Published
distributions of nanotube radii \cite{holt_fast_2006,hinds_aligned_2004,Dai_superlong_2011}
were used to generate the normal distribution of radii used in the
membranes, $r_{M}$, with mean $\left\langle r_{M}\right\rangle $
and standard deviation $\sigma_{M}$, used in our calculations. $Q_{M,0}$
is the total flow rate through a $1\,\mathsf{cm^{2}}$ membrane comprised
of frictionless nanotubes of radial distribution $r_{M}$. $b$ is
the slip length from Eq. \ref{eq:slipeq}. Values of $\mathcal{Q}=Q_{M}/Q_{M,0}>1$
do not comply with known physics of fluid flow. \label{tab:Table-1.-Data}}

\end{table}

For aquaporin, data is derived from published osmotic permeabilities
$p_{f}$ \cite{Preston1992} expressed in terms of hydrostatic pressure
drop across the channel, \[
\Delta P_{T}=\frac{\rho kT}{mp_{f}}Q,\]
 where $\rho$ is fluid density, $kT$ is the thermal energy, and
$m$ is the molecular mass. The end-effect pressure drop is given
by Eq. \ref{eq:endpressure}, with the channel radius chosen as the
average radius through the single-file portion of the channel.

Values used to produce Figures 2 and 3 and a summary of experimental
measurements are given in Table \ref{tab:Table-1.-Data}. In the figures,
the friction factor, $f$, is given in units of $\upmu\mathrm{m}^{-1}$.
The flow resistance, $R_{\alpha}=\Delta P_{\alpha}/Q$, where $\alpha=T,\mathcal{E}$
for total membrane or end-effects, respectively, and $\Delta P_{\alpha}$
is the pressure drop.

In Figure 1, the rate of energy dissipation was calculated from a
numerical simulation of the Navier-Stokes equations using an axially
symmetric geometry of a cylindrical channel between two large reservoirs
with a flow rate of $10^{3}\upmu\mathrm{m}^{3}/$s. A perceptually-balanced
color scale in MATLAB was chosen for visual accuracy to reduce luminance
variation artifacts of the standard rainbow colormap. In the simulations,
changing from slip to no-slip boundary conditions on the reservoir
walls had negligible effects on flow rates. Additionally, interactions
between neighboring channel openings were investigated by adding a
second nearby channel. At the separation for adjacent double-walled
carbon nanotubes and internal channel radius $1\:\mathrm{nm}$, such
as may occur in fabricated membranes \cite{holt_fast_2006}, no significant
interactions were observed, in agreement with theoretical predictions
\cite{wang_stokes_1994}.

\vspace{0.4in}

\noindent \textbf{Acknowledgements }Mark Johnson provided insight
into flows through biological pores. Mitra Hartmann and Alphonso Mondragon
provided editorial assistance. This work was funded by a generous
grant from the A. K. Barlow Foundation.

\pagebreak{}

\begin{onehalfspace}
\noindent \textbf{\large References}{\large \par}

\noindent \vspace{-0.9in}
\bibliographystyle{spbasic2}
\bibliography{endEffects}

\end{onehalfspace}

\end{document}